\documentstyle[apj,times,PSfig]{article}

\newcommand{\bfig}{\begin{minipage}{3.3in}\bigskip}
\newcommand{\efig}{\bigskip\end{minipage}}

\begin{document}

\title{Off-Axis Afterglow Emission from Jetted Gamma-Ray Bursts}

\author{Jonathan Granot$^1$, Alin Panaitescu$^2$, Pawan Kumar$^1$, and Stan Woosley$^3$}
\affil{$^1$ Institute for Advanced Study, Olden Lane, Princeton, NJ 08540}
\affil{$^2$ Dept. of Astrophysical Sciences, Princeton University, Princeton, NJ 08544}
\affil{$^3$ Dept. of Astronomy \& Astrophysics, University of California,
                 1156 High Street, Santa Cruz, CA 95064} 

\begin{abstract}

We calculate Gamma-Ray Burst (GRB) afterglow light-curves from a relativistic jet 
of initial opening angle $\theta_0$, as seen by observers at a wide range of viewing angles, 
$\theta_{\rm obs}$, from the jet axis. We describe three increasingly more realistic 
models and compare the resulting light-curves.  An observer at $\theta_{\rm obs}<\theta_0$ 
should see a light curve very similar to that for an on-axis observer.
An observer at $\theta_{\rm obs}>\theta_0$ should see a rising light curve at early times, 
the flux peaking when the jet Lorentz factor $\sim 1/\theta_{\rm obs}$. After this time the 
flux is not very different from that seen by an on-axis observer. 
A strong linear polarization ($\lesssim 40\%$) may occur near the peak in the light curve, 
and slowly decay with time.  
We show that if GRB jets have a universal energy, then orphan afterglows associated
with off-axis jets should be seen up to a constant $\theta_{\rm obs}$, therefore the
detection rate of orphan afterglows would be proportional to the true GRB rate.
We also discuss the proposed connection between supernova 1998bw and GRB 980425.

\end{abstract}


\section{Introduction}
\label{sec:intro}

Gamma-Ray Bursts (GRBs) are explosions which release roughly 10$^{51}$ erg
in the form of kinetic energy of highly relativistic
material (Frail et al. 2001, Panaitescu \& Kumar 2001)\footnote{
  Most of the information we have about GRB explosions is only for the so-called 
  long bursts, lasting more than a few seconds}. 
Many GRBs appear to be highly non-spherical explosions, as evidenced by a 
nearly-achromatic break in the light-curve (e.g. Harrison et al. 1999; Stanek et al. 1999).
%  \footnote{ There are alternate explanations for these breaks (e.g. Dai \& Lu 1999, 
%  Huang et al. 2000), but these models do not seem to be able to explain all available data}.
Highly relativistic jets are ``visible'' when our line of sight is within the jet 
aperture ($\theta_{\rm obs}<\theta_0$), otherwise, because of relativistic beaming 
of photons away from our line-of-sight, the object is too dim. As the jet 
decelerates, the relativistic beaming becomes less severe and the emission
from the jet becomes detectable to observers at larger viewing angles.

In this Letter we study the afterglow light-curves for off-axis locations 
($\theta_{\rm obs}>0$), focusing on observers lying outside of the initial jet opening 
angle ($\theta_{\rm obs}>\theta_0$).
Granot et al. (2001) have shown that the light curve seen by an observer located
within the initial jet aperture ($\theta_{\rm obs}<\theta_0$) is very similar to 
that for an on-axis observer ($\theta_{\rm obs}=0$). 
Dalal et al. (2002) and Rossi et al. (2002) have presented simple models to 
calculate the flux in this case. We reanalyze these models in \S2.1 and consider 
more realistic models in \S2.2 \& \S2.3. Moderski, Sikora and Bulik (2000) have calculated
off-axis light-curves with a more complex model, similar to that presented in \S2.2.

In \S3 we calculate the temporal evolution of the linear polarization for various
$\theta_{\rm obs}$. In \S4 we analyze the prospects of using the detection rate of orphan
afterglows to estimate the collimation of GRB jets. In \S5 we analyze the suggestion of 
Woosley, Eastman, \& Schmidt (1999) that a relativistic jet emanating from the SN explosion 
and pointing away from us could explain the observations.

%%%%%%%%%%%%%%%%%%%%%%%%%%%%%%%%%%%%%%%%%%%%%%%%%%%%%%%%%%%%%%%%
\section{Modeling the off axis emission}
\label{OAE}

In this section we calculate the afterglow light curves of jetted GRBs,
as seen by observers at different viewing angles, $\theta_{\rm obs}$,
w.r.t the symmetry axis of the jet. For simplicity, we
consider only a jet propagating into a homogeneous medium.
In order to improve our understanding of the underlying physics 
and in order to check how general the results are, we explore three different models
with an increasing level of complexity.

\subsection{Model 1: A Point Source at the Jet Axis}
\label{model1}

We begin with a simple model, where for $\theta_{\rm obs}=0$ the light curve follows the 
results of simple jet models (Rhoads 1999; Sari, Piran \& Halpern 1999,
hereafter R-SPH99), and for $\theta_{\rm obs}>0$ the light curves are calculated assuming
the emission is from a point source that moves along the jet axis.
%A similar model was used by Dalal et al. (2002), 
%however they concentrated on the bolometric luminosity, while we calculate the flux
%per unit frequency which is more useful for comparison with observations.
The on-axis light curve exhibits a jet break at (R-SPH99):
\begin{equation}
\label{t_jet}
t_{\rm jet}=6.2(1+z)(E_{52}/n_{0})^{1/3}(\theta_0/0.1)^{8/3}\ {\rm hr}\ ,
\end{equation}
where $E_{52}$ is the isotropic equivalent energy in units of $10^{52}$ erg,
$n_0$ is the ambient density in ${\rm cm}^{-3}$
and $z$ is the cosmological redshift of the
source. At $t<t_{\rm jet}$, $F_{\nu}(\theta_{\rm obs}=0)$
is taken from Sari, Piran and Narayan (1998), while
at $t>t_{\rm jet}$ the temporal scalings of the break frequencies and peak flux
change according to R-SPH99. 
The observed flux density from a point source is 
\begin{equation}
\label{point_source}
F_{\nu}
={L'_{\nu'}\over 4\pi d_{A}^2}\left({\nu\over \nu'}\right)^3
={(1+z)\over 4\pi d_{L}^2}{L'_{\nu'}\over\gamma^3(1-\beta\cos\theta)^3}\ ,
\end{equation}
where $L'_{\nu'}$ and $\nu'$ are the spectral luminosity and frequency in the
local rest frame of the jet, $d_{A}$ and $d_{L}$
are the angular and luminosity distances to the source, $\gamma=(1-\beta^2)^{-1/2}$
is the Lorentz factor of the source and $\theta$ is the angle between the direction
of motion of the source and the direction to the observer in the observer frame
(in our case $\theta=\theta_{\rm obs}$). Since
$t/t'\approx dt/dt'=\nu'/\nu=(1+z)\gamma(1-\beta\cos\theta)$,
where $t$ and $\nu$ are the observed time and frequency, we obtain that
\begin{equation}
\label{t_nu}
t_0/t_{\theta}=\nu_{\theta}/\nu_0=(1-\beta)/(1-\beta\cos\theta)
\equiv a \approx(1+\gamma^2\theta^2)^{-1}\ ,
\end{equation}
where $t_{\theta}$ and $\nu_{\theta}$ are the observed time and frequency for an
observer at $\theta_{\rm obs}=\theta$. One therefore obtains that
\begin{equation}
\label{F_nu}
F_{\nu}(\theta_{\rm obs},t)=a^3 F_{\nu/a}(0,at)\ ,
\end{equation}
where, for simplicity, we take $\gamma=\theta_0^{\, -1}[t_0/t_{\rm jet}]^{-3/8}$ at $t_0<t_{\rm jet}$
and $\gamma=\theta_0^{\, -1}[t_0/t_{\rm jet}]^{-1/2}$ at $t_0>t_{\rm jet}$.

The light curves obtained using equation \ref{F_nu} are shown by the dashed lines in
Figure \ref{fig1}. At first $\gamma\theta_{\rm obs}\gg 1$ and 
$a^3\approx(\gamma\theta_{\rm obs})^{-6}$ is the dominant term in equation \ref{F_nu},
giving a sharp rise in the light curve. 
Once $\gamma$ becomes $\lesssim\theta_{\rm obs}^{-1}$ the
flux begins to decay, asymptotically approaching the on-axis light curve. The light curve for
off-axis observers peaks when $\gamma\sim 1/\theta_{\rm obs}$.

\bfig
\plotone{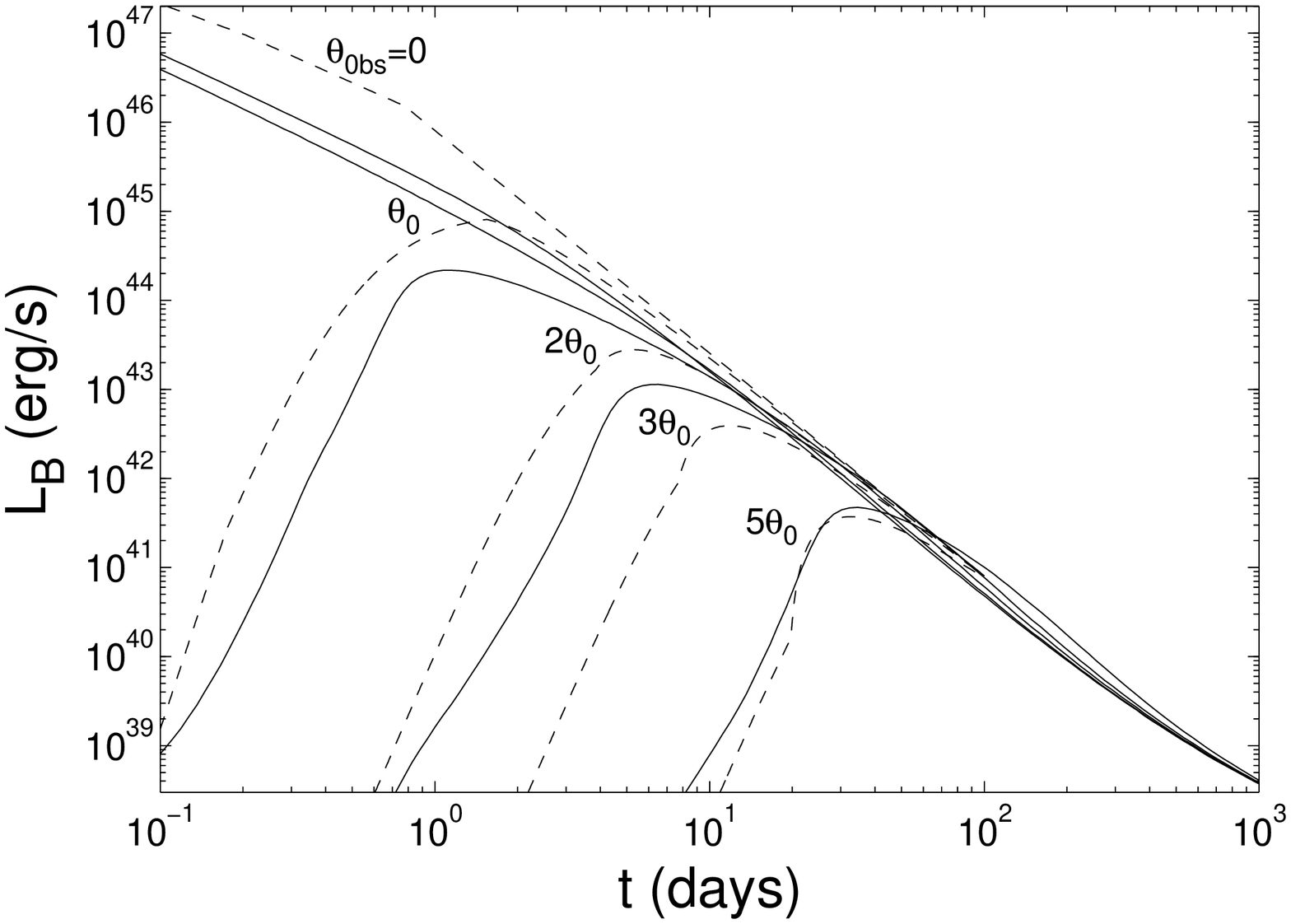}
\figcaption[]{\label{fig1}
B-band luminosity for models 1 (dashed lines) and 2 (solid lines),
for $\theta_0=5^\circ$, $\theta_{\rm obs}=(0,1,2,3,5)\theta_0$, $E_{52}=80$, $n_0=1$,
$p=2.5$, $\epsilon_B=0.01$, $\epsilon_e=0.1$, where $\epsilon_B$
($\epsilon_e$) is the fraction of the internal energy in the magnetic field
(electrons) and $p$ is the power law index of the electron energy distribution.
Model 1 is scaled down by a factor of 2.5, to help compare between the two models.  }
\efig

The main advantage of this model is that it is very simple, and nevertheless
gives reasonable results for $\theta_{\rm obs}\gtrsim 2\theta_0$.
Its main drawback is that it is not physical for
$\theta_{\rm obs}\lesssim\theta_0$ at $t\lesssim t_{\rm jet}$. This is because for
$\theta_{\rm obs}<\theta_0$ the observed radiation is initially dominated by emission from the
material within an angle of $1/\gamma<\theta_{\rm obs}$ around the line of sight, while in 
model 1 the emission is always only from along the jet axis,
and therefore the dominant contribution to the emission is missing, until the time when
$\gamma\sim 1/\theta_{\rm obs}$. This problem is overcome by our next model.

\subsection{Model 2: A Homogeneous Jet}
\label{model2}

This model is described in Kumar \& Panaitescu (2000), and here we briefly
point out its main features. The Lorentz factor and energy density per unit
solid angle are considered to be independent of angle $\theta$ 
within the jet aperture.
The decrease of the Lorentz factor of the jet with time is calculated
from the mass and energy conservation equations, and the sideway expansion
speed of the jet is taken to be the local sound speed.

The radiation calculation includes the synchrotron and inverse Compton processes,
and the synchrotron spectrum is taken to be piece-wise power-law with the usual
self-absorption, cooling and the synchrotron peak frequencies calculated from
the electron spectrum, magnetic field strength and the radiative loss of energy
for electrons. The observed flux is obtained by integrating the emissivity over
equal arrival time surface (e.g. Kumar \& Panaitescu 2000).

The light curves of model 2 are shown by the solid lines in figure \ref{fig1}.
The flux density in the decaying stage is slightly higher for larger viewing
angles $\theta_{\rm obs}$. This effect occurs since at this late stage the whole
jet is visible, and for larger $\theta_{\rm obs}$ the radiation from a given radius
arrives at the observer at a latter time, on average. Therefore, for a given 
observed time, larger $\theta_{\rm obs}$ are dominated by emission from smaller radii,
resulting in a larger flux density.
At a few hundred days, the light curves begin
to flatten due to the transition to the non-relativistic regime.

The light curves for $\theta_{\rm obs}\lesssim\theta_0$ are very different from model 1
(and more realistic). Furthermore, the light curves for $\theta_{\rm obs}\le\theta_0$ are
very similar to $\theta_{\rm obs}=0$ in this model. Since the jet is homogeneous,
the ratio of the observed flux for $\theta_{\rm obs}<\theta_0$ and $\theta_{\rm obs}=0$,
may be approximated by the ratio of the areas within the jet, that are within an angle of
$1/\gamma$ around the directions to these two observers (which never decreases below $1/2$).

We notice that the lights curve of model 1 for $\theta_{\rm obs}/\theta_0=1,\, 2$ are much 
closer to the light curves of model 2 for $\theta_{\rm obs}/\theta_0=2,\, 3$, respectively, 
than to the light curves for the same viewing angles. This is so because the emission 
for an observer outside the jet opening angle is dominated by the point in the jet 
closest to the observer. Therefore, model 1 will become much more realistic (and just
as simple) if one would use $\theta=\max(0,\theta_{\rm obs}-\theta_0)$, rather than 
$\theta=\theta_{\rm obs}$, in equations \ref{point_source} and \ref{t_nu}.

The main advantage of model 2 is that it provides realistic light curves in a very
reasonable computational time, making it very convenient for performing
detailed fits to observations (e.g. Panaitescu \& Kumar 2001). Its main drawback is a
relatively simple treatment of the dynamics, which causes some differences in the light
curves, compared to our next model.

\subsection{Model 3: 2D Hydrodynamical Simulation}
\label{model3}

This model is described in Granot et al. (2001). The jet dynamics
are determined by a 2D hydrodynamical simulation, with initial conditions 
of a wedge taken from the spherical self similar Blandford-McKee (1976) solution.
The light curves for observers at different $\theta_{\rm obs}$ are calculated considering
the contribution from all the shocked region, and taking into account the relevant
relativistic transformations of the radiation field, and the different photon arrival
times to the different observers.

\bfig
\plotone{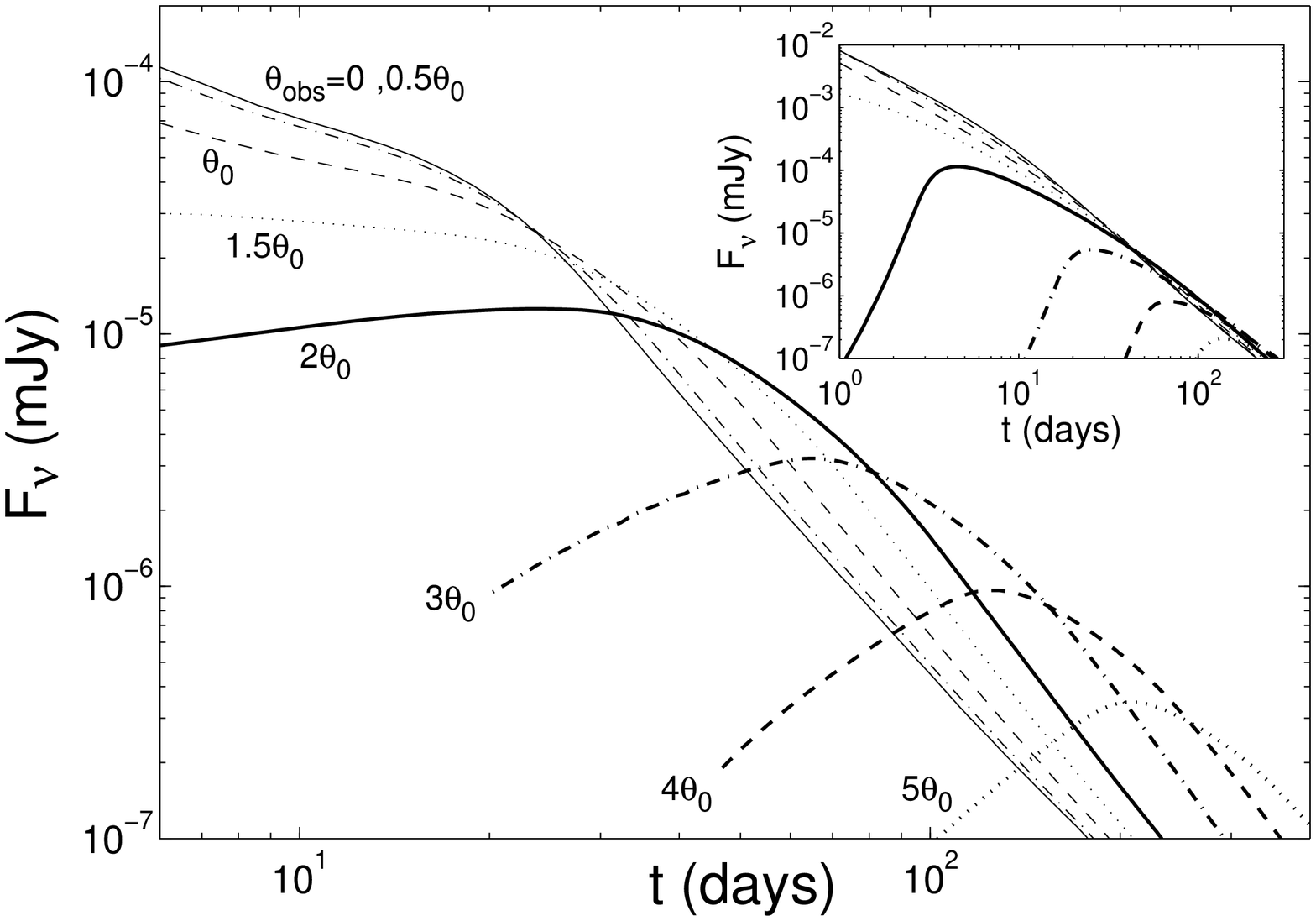}
\figcaption{\label{fig2}
Light curves of model 3, for $\theta_0=0.2$, $E_{52}=n_0=z=1$, $p=2.5$, $\epsilon_e=0.1$,
$\epsilon_B=0.01$, and $\nu=5\cdot 10^{14}$ Hz. The inset shows the same light curves for model 2,
where the same traces correspond to the same viewing angles $\theta_{\rm obs}$.}
\efig

Figure \ref{fig2} shows the light curves of models 3, while the inset provides the
light curves of model 2, for the same set of parameters. In model 3, the peak of
the light curves for $\theta_{\rm obs}>\theta_0$ is flatter compared to model 2, and
is obtained at a somewhat latter time. The rise before the peak is not as sharp as in 
models 1 or 2, since in model 3 there is some material at the sides of the jet with a 
moderate Lorentz factor (Granot et al. 2001; Piran \& Granot 2001). The emission from 
this slower material tends to dominate the observed flux at early times for observers
at $\theta_{\rm obs}>\theta_0$, resulting in a gentler rise before
the peak. The light curves for $\theta_{\rm obs}>\theta_0$ peak at a later time compared 
to model 2, and the flux during the decay stage grows faster with $\theta_{\rm obs}$, 
since in model 3 the curvature of the shock front is larger and the emission occurs within 
a shell of finite width, resulting in a larger photon arrival time, and implying that 
smaller radii contribute to a given observer time. The light-curves
for model 2 \& 3 are quantitatively similar for $\theta_{\rm obs}<\theta_0$.

The main advantage of this model is a reliable and rigorous treatment of the jet dynamics,
which provides insight on the behavior of the jet and the corresponding light curves. 
Its main drawback is the long computational time it requires.

%%%%%%%%%%%%%%%%%%%%%%%%%%%%%%%%%%%%%%%%%%%%%%%%%%%%%%%%%%%%%%%%%%%%
\section{Linear Polarization}
\label{polarization}

While the afterglow emission from a spherical outflow is expected to exhibit 
little or no linear polarization, as the polarization from the different parts 
of the afterglow image cancel out, a jetted outflow breaks the circular symmetry 
of the afterglow image and may exhibit a polarization of up to $\lesssim 20\%$
(Ghisellini \& Lazatti 1999; Sari 1999). One might therefore expect an even
larger polarization for an observer at $\theta_{\rm obs}>\theta_0$. 

An isotropic magnetic field configuration in the local rest frame will produce
no linear polarization. However, as the magnetic field is most likely produced at the 
shock itself, 
%the plane of the shock front defines a preferred direction, and 
one might expect the magnetic field perpendicular ($B_\perp$) and parallel
($B_\parallel$) to the shock direction, to have different magnitudes 
(Gruzinov 1999; Sari 1999). We calculate the
linear polarization for model 2 following Ghisellini \& Lazatti (1999)\footnote{Our 
calculation is different from Ghisellini \& Lazatti in that we include 
lateral expansion of the jet 
and integration over equal photon arrival time surfaces, so our results for 
$\theta_{\rm obs}<\theta_0$ are different.} 
and using their notations. They assume the magnetic field is strictly in the plane of 
the shock ($B=B_{\perp}$).

\bfig
\plotone{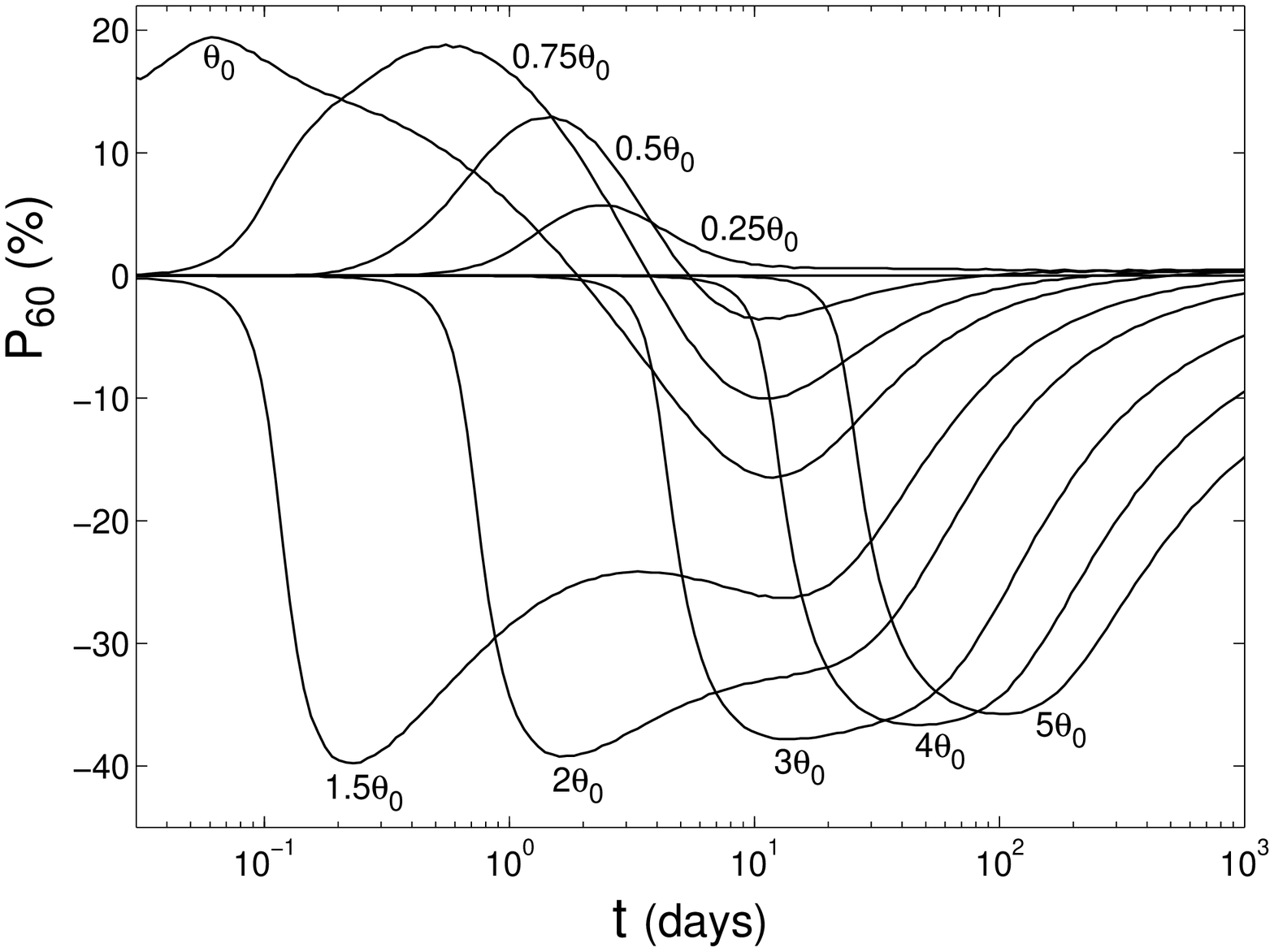}
\figcaption{\label{fig3}
The linear polarization for model 2 for several viewing angle
and for the same parameters as in Figure \ref{fig1}. }
\efig

Figure \ref{fig3} shows the polarization as a function of time for different 
$\theta_{\rm obs}$ in terms of $P_{60}$. For $P_{60}<0$ the polarization is along the 
plane containing the line of sight and the jet axis, wile for $P_{60}>0$ it is rotated 
by $90^\circ$ (for $\langle B_{\perp}\rangle < 2\langle B_{\parallel}\rangle$ this is
reversed, e.g. Sari 1999). A more isotropic magnetic field configuration would result in a
smaller degree of polarization, so the value of the polarization in Figure \ref{fig1} 
($\lesssim 40\%$) may be viewed as a rough upper limit. 
For $0.3 \lesssim \theta_{\rm obs}/\theta_0 \lesssim 1.1$ the polarization vanishes
and reappears rotated by $90^\circ$ around $t_{\rm jet}$. This behavior may occur 
again at a later time, but the subsequent polarization is very low. 
For $1.1 \lesssim \theta_{\rm obs}/\theta_0 \lesssim 1.6$ the polarization has two peaks, the 
first higher than the second. For $\theta_{\rm obs}/\theta_0\gtrsim 1.1$
%observers outside the initial opening angle of the jet,
the polarization is largest near the peak in the light curve, and decreases quite 
slowly with time, while the peak polarization shows a very weak dependence on 
$\theta_{\rm obs}$, and is about a factor of $2$ larger than for $\theta_{\rm obs}=\theta_0$. 

%%%%%%%%%%%%%%%%%%%%%%%%%%%%%%%%%%%%%%%%%%%%%%%%%%%%%%%%%%%%%%%%%%%%
\section{Orphan Afterglows}
\label{OA}

If GRB jets have well defined edges, the prompt gamma-ray flux drops very sharply
outside the opening of the jet, and the prompt burst will be very hard to detect from
$\theta_{\rm obs}>\theta_0$. On the other hand, the afterglow emission may be detected
out to $\theta_{\rm det}\sim\ {\rm a\ few}\ \theta_0$, where the exact value of 
$\theta_{\rm det}$ depends on the jet parameters (including its redshift), the observed 
band and the limiting flux for detection. Jetted GRBs with 
$\theta_0<\theta_{\rm obs}<\theta_{\rm det}$ are expected 
to be orphan afterglows (i.e. detectable in the optical but not in gamma-rays). 

It has been argued by Dalal et al. (2002) that $\theta_{\rm det}/\theta_0\approx{\rm const}$
for $\theta_0\ll 1$, so that the detection rate of orphan afterglows 
$\dot{\rm N}_{\rm orph}^{\rm det}$ (associated with off-axis jets) will be a constant 
[namely $(\theta_{\rm det}/\theta_0)^2$] times the GRB detection rate 
$\dot{\rm N}_{\rm GRB}^{\rm det}$, and thereby a comparison between these two rates 
will not constrain $\theta_0$ or the true rate of GRBs $\dot{\rm N}_{\rm GRB}^{\rm true}$. 
This result was obtained assuming a constant flux, $F_\nu(t_{\rm jet})$, at $t_{\rm jet}$ 
for $\theta_{\rm obs}=0$. However, afterglow observations 
suggest that the total energy in the jet, $E_{\rm jet}$, is roughly constant 
(Frail et al. 2001, Panaitescu \& Kumar 2001, Piran et al. 2001) while $F_{\nu}(t_{\rm jet})$ 
varies over a wider range. In fact, for $E_{\rm jet}={\rm const}$, simple jet models (R-SPH99) 
predict that the hydrodynamical evolution of the jet (and therefore the light curves for 
all $\theta_{\rm obs}$) becomes independent of $\theta_0$ once the jet enters the phase of 
exponential lateral expansion with radius. This corresponds to $t>t_{\rm jet}$ for 
$\theta_{\rm obs}<\theta_0$, and to $t\gtrsim(\theta_{\rm obs}/\theta_0)^2t_{\rm jet}$ 
for $\theta_{\rm obs}>\theta_0$, which includes the time around or after the peak in 
the light curve. 
%This may be understood since once the jet enters the phase of exponential lateral 
%expansion with radius, the system 'forgets' the initial conditions (i.e. $\theta_0$) 
%and has knowledge only of the total energy in the jet. 
This implies that for $E_{\rm jet}={\rm const}$, we have $\theta_{\rm det}=
{\rm const}\equiv\theta_{\rm det,0}$ (rather than $\theta_{\rm det}/\theta_0={\rm const}$)
for $\theta_0<\theta_{\rm det}$. For $\theta_0\gtrsim\theta_{\rm det,0}$
naturally $\theta_{\rm det}$ is larger $\theta_0$ if the afterglow is
detectable from $\theta_{\rm obs}<\theta_0$, and the solid angle between $\theta_0$ 
and $\theta_{\rm det}$, $\Omega_{\rm orph}=\cos\theta_0-\cos\theta_{\rm det}
\approx(\theta_{\rm det}^2-\theta_0^2)/2$ (in which we have detectable orphan afterglows) 
remains approximately constant. Furthermore, the distribution of $\theta_0$ 
inferred from observations (Frail et al. 2001, Panaitescu \& Kumar 2001) is 
sharply peaked at low $\theta_0$ ($\sim 2-3^\circ$). This suggests 
that most of the orphan afterglows that would be detected should have 
$\theta_0\sim 3^\circ$. For a reasonable limiting magnitude for detection, this 
implies $\theta_{\rm det}\gg\theta_0$ in most cases, and therefore $\theta_{\rm det}\approx
\theta_{\rm det,0}$.
For example, for model 2 with $E_{\rm jet}=5\cdot 10^{50}$ ergs (assuming a double sided 
jet), $\epsilon_e=0.1$, $\epsilon_B=0.01$, $p=2.5$, $n_0=z=1$ and a limiting magnitude 
for detection of $R=24$ we obtain $12.4^\circ<\theta_{\rm det}<23^\circ$ and 
$0.023<\Omega_{\rm orph}/4\pi<0.045$ for $2^\circ<\theta_0<15^\circ$.  
If indeed $\Omega_{\rm orph}\approx {\rm const}$, then $\dot{\rm N}_{\rm orph}^{\rm det}$ 
should provide a good estimate of the true GRB rate, 
$\dot{\rm N}_{\rm GRB}^{\rm true}=(4\pi/\Omega_{\rm orph})\dot{\rm N}_{\rm orph}^{\rm det}$. 
The average beaming fraction 
$f_b=\langle 1-\cos\theta_0\rangle\approx\langle \theta_0^2\rangle/2$ is given by 
$f_b=\dot{\rm N}_{\rm GRB}^{\rm det}/\dot{\rm N}_{\rm GRB}^{\rm true}
=(\Omega_{\rm orph}/4\pi)\dot{\rm N}_{\rm GRB}^{\rm det}/\dot{\rm N}_{\rm orph}^{\rm det}$.

%%%%%%%%%%%%%%%%%%%%%%%%%%%%%%%%%%%%%%%%%%%%%%%%%%%%%%%%%%%%%
\section{GRB 980425 / SN 1998bw}

On April 25, 1998, a Gamma-Ray Burst was detected by Beppo SAX and CGRO. The burst 
consisted of a single wide peak of duration ~30 s, peak flux in 24-1820 keV band of
$3\cdot 10^{-7}$ erg cm$^{-2}$ s$^{-1}$, and fluence of $4.4\cdot 10^{-6}$ erg cm$^{-2}$
(Soffitta et al. 1998, Kippen et al. 1998). The burst had no detectable emission 
above 300 keV. The burst spectrum was a broken power-law with break at 148$\pm$33 keV, 
and the high energy power-law photon index of $-3.8\pm 0.7$ (see Galama et al. 1998).
These values are not unusual for GRBs.

A bright Type Ic supernova, SN 1998bw, located at $z=0.0085$, was detected within 8 arc
minutes of GRB 980425. From the extrapolation of optical light curves Galama et al.
(1998) suggested that the SN went off within a day of the GRB, thereby implying a
possible connection between the two events. The probability of this association is
strengthened by the uniquely peculiar light curve and spectrum of the SN (e.g., Patat
et al 2001). Early on, Woosley et al. (1999) gave arguments why SN 1998bw might be a
SN exploded by a jet and therefore possibly associated with a GRB. This would arise,
for instance, in the collapsar model (Woosley 1993; MacFadyen and Woosley 1999; 
Paczy\'nski 1998).

If indeed the two events are associated, then the total isotropic equivalent of energy
release in $\gamma$-rays for GRB 980425 is $E_{\gamma,iso}=8.5\cdot 10^{47}$ erg, or
a factor of $\sim 10^4$ smaller than the energy for an average cosmological GRB.
This could explained in two ways.

\subsection{Sharp Edged, Homogeneous Jet Seen at $\theta_{\rm obs}>\theta_0$}

If GRB 980425 was viewed at $\theta_{\rm obs}>\theta_0$ it might explain 
its small $E_{\gamma,iso}$. For a GRB arising from a jet with $\gamma$ independent of
$\theta$ and sharp edges, the observed energy falls off rapidly for $\theta_{\rm obs}>\theta_0$,
in fact as $b^6$ where $b\equiv\gamma(\theta_{\rm obs}-\theta_0)$. Moreover, for an observer 
at $\theta_{\rm obs}>\theta_0$ the peak of the spectrum is lower by a factor $b^2$, and the 
burst duration longer by the same factor, compared to an observer at $\theta_{\rm obs}<\theta_0$. 
Therefore the low $E_{\gamma,iso}$ of 980425 implies $b^6\sim 10^4$ and $\theta_{\rm obs}\sim 
\theta_0 + 5^{\circ} (\gamma/50)^{-1}$. If GRB 980425 were at a cosmological distance and
seen from $\theta_{\rm obs}<\theta_0$, the peak of the spectrum and the burst duration would be 
$\sim 1[3/(1+z)]$ MeV and 4$[(1+z)/3]$ s, respectively.

A second constraint is set by the condition that the optical afterglow is 
dimmer than SN 1998bw, which had a luminosity of $2\times 10^{42}$ erg at 1 day,
rose to $10^{43}$ erg at $14-19$ days (Galama et al. 1998), then decayed at
$\sim 0.017$ mag/day (Patat et al. 2001). To compare it with the afterglow luminosity,
we shall use the typical properties of the afterglows whose optical light-curves exhibited 
breaks: $i)$ average jet break $t_{\rm jet} \sim 0.5$ days in their rest-frame,
$ii)$ average optical luminosity flux $L(\theta_{\rm obs}=0,t_{\rm jet}) \sim 2\times 10^{45}$
erg at $t_{\rm jet}$, $iii)$ $L \propto t^{-\alpha}$ with $\alpha \sim 2$ at $t>t_{\rm jet}$, 
$iv)$ $L_\nu \propto \nu^{-\beta}$ with $\beta \sim 1$, at optical frequencies.

 Using Model 1 described in section \S2.1, it can be shown that for an observer at 
$\theta_{\rm obs} \gtrsim 2\theta_0$ the time and optical luminosity at the light-curve peak are
\begin{equation}
 t_{peak} \simeq \left[ 5 + 2\ln \left( \frac{\theta_{\rm obs}}{\theta_0} - 1 \right) \right]
               \left( \frac{\theta_{\rm obs}}{\theta_0} - 1 \right)^2\; t_{\rm jet} \;,
\end{equation}
\begin{equation}
 L_{peak} (\theta_{\rm obs}) \simeq 2^{-(\beta+3)}  \left( \frac{\theta_{\rm obs}}{\theta_0} - 1
           \right)^{-2\alpha} L (0,t_{\rm jet}) \;.
\end{equation}
From these equations it can be shown that for $\theta_{\rm obs} \gtrsim 3\,\theta_0$ the peak 
afterglow luminosity $L_{peak}(t_{peak})$ is a factor $\sim 3$ lower than $L(t)$ of SN 1998bw.
During the decay phase, the afterglow luminosity remains below that of the SN until about
400 days, when they become comparable. We note that Patat et al. (2001) report a flattening
of the SN 1998bw decay after 375 day, to a dimming rate of $\sim 0.009$ mag/day, which they
interpret as the settling in of the $^{56}$Co decay or the existence of other sources.

 Together with the above constraint, $\theta_{\rm obs}\sim \theta_0 + 5^{\circ} (\gamma/50)^{-1}$, 
imposed by the fluence of GRB 980425, the condition that the afterglow emission does not 
exceed that of the SN 1998bw leads to $\theta_0 \lesssim 3^{\circ} (\gamma/50)^{-1}$ and 
$3\theta_0 \lesssim \theta_{\rm obs} \lesssim 8^{\circ} (\gamma/50)^{-1}$.

\subsection{Inhomogeneous Jet Seen Off-Axis}

Another possibility is that the jet does not have sharp
edges, but wings of lower energy and Lorentz factor that extend to large $\theta$.
Such a picture of the jet was suggested by Woosley et al. (1999) and is consistent
with the relativistic studies of the collapsar model by Zhang, Woosley, \& MacFadyen
(2002). GRB 980425 would be then be produced by material with $\gamma \sim 10$ moving
in our direction.

 When seen on-axis, an afterglow with the above properties has $L \sim 5\times 10^{44}$ 
erg at 1 day and $L \sim 2\times 10^{42}$ erg at 16 days, i.e. a "typical" afterglow would 
be 200 times brighter than SN 1998bw when its first observation was made and several times 
dimmer when SN 1998bw peaked.  All other parameters remaining the same, 
an afterglow emission dimmer than that of the SN 1998bw at 1 day requires an energy per
solid angle $\epsilon_{\rm jet}(\theta_{\rm obs})$ in the direction toward the observer 
satisfying $\epsilon (\theta_{\rm obs}) \theta_{\rm obs}^2 \sim (200)^{-3/(p+3)} E_{\rm jet} 
\sim 2\cdot 10^{49}$ erg. As in the previous subsection, the observer location satisfies
$\theta_{\rm obs} \gtrsim 3\theta_0$ so that the optical emission of the main jet of
opening $\theta_0$ is below the supernova light-curve.

\section{Discussion}
\label{discussion}

We have presented the calculation of light-curves from a relativistic jet
for an arbitrary location of the observer; much of the work in this letter is for
an observer located outside the initial jet opening, $\theta_{\rm obs}>\theta_0$.
We have considered three different jet models of increasing
sophistication; the simplest being a point source moving along the jet axis (\S2.1),
and the most sophisticated is 2D hydrodynamical simulation (\S2.3). The basic
qualitative features of the light-curves are similar in all three models, for
$\theta_{\rm obs}>\theta_0$. Moreover, the uniform jet model (model 2, \S2.2) is 
in rough quantitative agreement with the hydro-model.

We find that "orphan" optical afterglows associated with off-axis jets can be observed 
up to a constant $\theta_{\rm obs}$, rather than a constant $\theta_{\rm obs}/\theta_0$ as 
suggested by Dalal et al. (2002), if one assumes a constant energy in the jet, rather 
than a constant flux at the time of the jet break for an on-axis observer.
This implies that future surveys for orphan afterglows may provide valuable data
for the the distribution of jet opening angles $\theta_0$ and the true event rate of GRBs.
The orphan optical events discussed here can be identified from the initial rise during 
which the spectral slope is typically $\beta > 0$, followed by a decay, on a time scale of 
$\sim 1-30$ days, and may show a large degree of linear polarization ($\lesssim 40\%$).
The detection of such orphan afterglows may provide a new line of evidence in favor of 
jetted outflows in GRBs. Recently Huang, Dai and Lu (2001) have considered another scenario 
(failed GRBs) for producing orphan afterglows; this would increase the detection rate of 
orphan afterglows. A good monitoring of optical transients may help distinguish failed GRBs
from jets seen at $\theta_{\rm obs} > \theta_0$, and improve our understanding of them.

\acknowledgements

We thank Mark Miller for generating the data for model 3.
This research was supported by grants NSF PHY 99-79985 and MCA93S025
(computational support), NSF grant PHY-0070928 (JG), Lyman Spitzer, Jr.
Fellowship (AP), NAG5-8128 and MIT-292701 (SW).

\end{document}